\documentclass{emulateapj}
\usepackage{graphicx}
\usepackage[usenames,dvips]{color}
\def\arcsec{$\,^{\prime\prime}$~}
\def\arcmin{$\,^\prime$~}

\newcommand{\be}{\begin{equation}}
\newcommand{\bel}[1]{\begin{equation}\label{eq:#1}}
\newcommand{\ee}{\end{equation}}
\newcommand{\bd}{\begin{displaymath}} 
\newcommand{\ed}{\end{displaymath}}   
\newcommand{\bea}{\begin{eqnarray}}
\newcommand{\beal}[1]{\begin{eqnarray}\label{eq:#1}}
\newcommand{\eea}{\end{eqnarray}}

\newcommand{\eqref}[1]{\ref{eq:#1}}

\newcommand{\lsim }{{\lower0.8ex\hbox{$\buildrel <\over\sim$}}}
\newcommand{\gsim }{{\lower0.8ex\hbox{$\buildrel >\over\sim$}}}

\def\Chandra{${\it Chandra}$\ }

\def\simge{\mathrel{%
   \rlap{\raise 0.511ex \hbox{$>$}}{\lower 0.511ex \hbox{$\sim$}}}}
\def\simle{\mathrel{
   \rlap{\raise 0.511ex \hbox{$<$}}{\lower 0.511ex \hbox{$\sim$}}}}

\newcommand{\Msun}{\ifmmode {M_{\odot}}\else${M_{\odot}}$\fi}
\newcommand{\Lsun}{\ifmmode {L_{\odot}}\else${L_{\odot}}$\fi}
\newcommand{\Rsun}{\ifmmode {R_{\odot}}\else${R_{\odot}}$\fi}

\shorttitle{The Unusual X-ray Binaries of the Globular Cluster NGC 6652}
\shortauthors{Coomber et al.}

\begin{document}
\title{The Unusual X-ray Binaries of the Globular Cluster NGC 6652}  

\author{G.~Coomber\altaffilmark{1}, C.~O. Heinke\altaffilmark{1,2}, H.~N.~Cohn\altaffilmark{3}, P.~M.~Lugger\altaffilmark{3}, J.~E. Grindlay\altaffilmark{4}}

\altaffiltext{1}{University of Alberta, Dept. of Physics, 11322-89 Avenue, Edmonton AB T6G 2G7, Canada}

\altaffiltext{2}{Ingenuity New Faculty; heinke@ualberta.ca}

\altaffiltext{3}{Dept. of Astronomy, Indiana University, 727 East 3rd St., Bloomington IN 47405, USA }

\altaffiltext{4}{Harvard-Smithsonian Center for Astrophysics, 60 Garden Street, Cambridge MA 02138, USA}


\begin{abstract}
Our 5 ks \Chandra\ ACIS-S observation of the globular cluster NGC 6652 detected 7 X-ray sources, 3 of which are previously unidentified.  This cluster hosts a well-known bright low-mass X-ray binary, source A (or XB 1832-330).  
Source B shows unusual rapid flaring variability, with an average $L_X$(0.5-10 keV) $\sim2\times10^{34}$ ergs/s, but with minutes-long flares up to $L_X=9\times10^{34}$ ergs/s.  Its spectrum can be fit by an  absorbed power-law of photon index  $\Gamma\sim1.24$, and hardens as the countrate decreases. This suggests that part or all of the variation might be due to obscuration by the rim of a highly inclined accretion disk.  
Sources C and D, with $L_X \sim 10^{33}$ ergs/s, have soft and unusual spectra.
Source C requires a very soft component, with a spectrum peaking at 0.5 keV, which might be the hot polar cap of a magnetically accreting polar cataclysmic variable.  
 Source D shows a soft spectrum (fit by a power-law of photon index $\sim2.3$) with marginal evidence for an emission line around 1 keV; its nature is unclear.   
The faint new sources E, F, and G have luminosities of 1-2$\times 10^{32}$ ergs/s, if associated with the cluster (which is likely).  E and F have relatively hard spectra (consistent with power-laws with photon index $\sim$1.5).  G lacks soft photons, suggesting absorption with $N_H>10^{22}$ cm$^{-2}$.

\end{abstract}


\maketitle

\section{Introduction}\label{s:intro}

The high-density environments of globular clusters were suggested to be causes of X-ray binary production early in X-ray astronomy \citep{Katz75,Clark75}.  Several types of X-ray sources have now been identified in Galactic globular clusters \citep{Verbunt06}.  Fifteen luminous low-mass X-ray binaries (LMXBs) reach $L_X>10^{35}$ ergs/s, some only during short outbursts \citep{Sidoli01,Verbunt06,Heinke10,Pooley10}.  The low-luminosity X-ray sources $L_X=10^{29}$ - $10^{34}$ ergs/s include several types of systems such as quiescent LMXBs (containing neutron stars between accretion episodes), millisecond radio pulsars, cataclysmic variables (CVs), and magnetically active binaries \citep{Verbunt06}. Quiescent LMXBs containing neutron stars (qLMXBs) often show soft thermal spectra dominated by blackbody-like emission \citep{Rutledge02a,Heinke03d}, though a spectrally harder component is often present and sometimes dominant \citep{Campana02,Jonker04,Wijnands05a}. The origin of the soft component is generally thought to be thermal X-ray emission from the heated NS surface, modified by the star's hydrogen atmosphere \citep{Zavlin96,Brown98}. The harder power-law component is not yet well understood, but may be produced by continuing accretion or possibly pulsar activity \citep{Campana98a}. 

The globular cluster NGC 6652 contains one luminous ($L_X>10^{36}$) LMXB, XB 1832-330 \citep{Predehl91} and several lower-luminosity X-ray sources. This core-collapsed \citep{Noyola06} cluster is $\sim$11.7 Gyr old \citep{Chaboyer00},  9.0$\pm$0.4 kpc from the sun and suffers extinction  with equivalent hydrogen column density $N_H$ = 5.0 x $10^{20}$ cm$^{-2}$ \citep{Harris96}.  The three fainter sources detected by \citet{Heinke01} in a 1.6 ks {\it Chandra} HRC-I observation are not well understood. The brightest of the three low-$L_X$ sources (source B) was seen at luminosities of a few $10^{33}$ ergs/s while its optical counterpart lies on the main sequence (1 magnitude below the turnoff) in a $V-I$ optical color-magnitude diagram, suggesting a qLMXB nature \citep{Heinke01}.  However, the optical counterpart has also been observed to show blue $U$-$V$ colors \citep{Deutsch98a}, and strong variability with a (possible) 43.6-minute period \citep{Deutsch00}, which, if it is the true period, would exclude a main-sequence companion.  Source C has a very blue optical counterpart, which indicates a bright disk and a relatively high rate of mass transfer.  Combining this with its relatively low $L_X$, the efficiency of energy extraction is inferred to be relatively low, suggesting a CV rather than an LMXB \citep{Heinke01}.  

In this paper we present a 2008 \Chandra\ ACIS-S observation of NGC 6652's X-ray sources. We also consult 2000 \Chandra\ HRC-I data and 1994 ROSAT HRI data for long-term variability information.

\section{Data Reduction}\label{s:obs}

We observed NGC 6652 on June 9, 2008 for 5.6 ks with \Chandra's ACIS-S detector in a 1/4 subarray mode.  The data were reduced using the CIAO version 4.3 software\footnote{Available at http://cxc.harvard.edu/ciao/}. We created a new bad pixel file with the \texttt{acis\_run\_hotpix} script. The level 1 event files were reprocessed by calibrating for charge-transfer inefficiency on the detector and time-dependent gain adjustments. The data was then filtered for grade and status bits according to the standard CIAO Science Threads\footnote{http://cxc.harvard.edu/ciao/threads/all.html}. A  background light curve shows no evidence of flaring. 

The CIAO WAVDETECT program was run using the energy range 0.3-7.0 keV. 
We find 7 sources, all of which are located within (or at) the cluster half-mass radius \textit{$r_h$}, 0\farcm65 \citep{Harris96}.  Three faint sources (E, F and G) were not visible in the HRC observation of NGC 6652 by  \citet{Heinke01}.
The positions of sources B through G, shown in Table 1, were computed using WAVDETECT centroiding. Due to the high flux of photons from source A, pileup effects 
produce a characteristic ``donut hole'' at the source location and a prominent readout streak, which prevents WAVDETECT from accurately computing the position of source A. We therefore estimated the location of source A by matching a symmetric circle to the ``donut hole'', by eye. Running WAVDETECT in the energy bands 0.3-2.0 keV, 2.0-5.0 keV, and 5.0-7.0 keV did not locate additional sources. The positions of sources A, B, C and D are 
consistent with those from the 2001 \textit{Chandra} observation. We show a smoothed ACIS image of NGC 6652 in Fig. 1. Using the log N-log S relationship of  \citet{Giacconi01}, we calculated the expected number of active galactic nuclei observed within one half-mass radius of the cluster to be 0.14, indicating that all detected X-ray sources are likely to be members of the cluster. 

Due to pileup effects, the data from the source region of A is not reliable.  Pileup occurs when two or more photons are recorded by the CCD as a single higher energy event, which results in distortion of the observed source spectrum and an underestimated count rate \citep{Davis01}.  The high count rate also produces a readout streak, as photons land on the detector during the short frame readout time.  By extracting a spectrum from the readout streak, we are able to obtain spectral data from source A without having to model the pileup. Using \texttt{dmextract} we extract a source spectrum from 
 the readout streak, and a background spectrum from surrounding source-free regions of the observation. We then create new RMF and ARF files using the tools \texttt{mkacisrmf} and \texttt{mkarf} respectively, and correct the streak exposure time to 110.4 s to account for the fact that the spectrum was extracted from the ACIS readout streak.

For sources B, C and D, we extracted 0-10 keV spectra using \texttt{specextract} from circular regions, of radius 2'' for B and 1.4'' for C and D, located at the source positions produced by WAVDETECT.  We produced spectra, response files and backgrounds following the appropriate CIAO thread\footnote{http://cxc.harvard.edu/ciao/threads/psextract/}, correcting the response files for the fraction of the flux contained within the extraction aperture, group the spectra, and fit them within XSPEC\footnote{http://heasarc.gsfc.nasa.gov/docs/xanadu/xspec/}.
Count rates were calculated for sources E, F and G directly from the level 2 event file using the \texttt{dmstat} tool, since these sources are too faint to allow spectral analysis. The count rates were then converted into unabsorbed fluxes with PIMMS\footnote{http://asc.harvard.edu/toolkit/pimms.jsp}, by assuming that the faint source spectra can be modeled by a power law of photon index $\Gamma$ = 1.4, typical for globular cluster sources at this $L_X$ \citep{Heinke05a}, and consistent with their hardness ratios (below).

To study long term variability of the sources, we re-analyze the archival HRC-I data from a 2000 observation of NGC 6652, 
 filtering the data for grade and status. We extract lightcurves for sources A, B, C, and D with the CIAO tool \texttt{dmextract} and estimate source luminosities with PIMMS.  We extract counts and lightcurves from source regions of radius 2'' centered at the WAVDETECT coordinates obtained from our ACIS observation. 

\begin{figure}
\figurenum{1}
\includegraphics[angle=0,scale=.4]{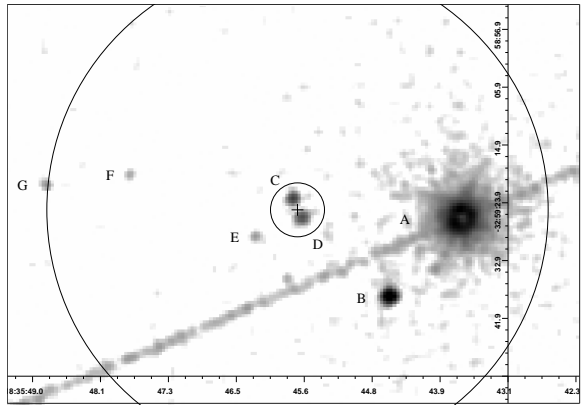}
\caption{\Chandra\ ACIS-S image of NGC 6652. The data is smoothed with a gaussian kernel radius of 1\arcsec. The 7 detected sources are labeled from A to G. Both the core radius (0.07\arcmin) and the half mass radius (0.65\arcmin) are indicated.} 
\end{figure}

\section{X-ray Analysis}\label{s:spec}


{\bf 3.1. Source A}

The 462 events extracted from source A's readout streak were grouped into 9 bins, with 50 counts per bin to maximize spectral resolution while minimizing the error in the normalized counts in each bin (Fig. 2). We ignore events above 8 keV, as the ACIS response is poorly understood at these energies.
The spectrum is well fit by an absorbed power-law model with a photon index of $\Gamma$ = $1.7^{+0.3}_{-0.2}$ (Table 2). 
The $N_H$ value of $2.7^{+0.1}_{-0.1}\times 10^{21}$ cm$^{-2}$ exceeds the accepted cluster value of 5$\times 10^{20}$ cm$^{-2}$ \citep{Harris96}. We derive an unabsorbed luminosity for source A of $L_X$(0.5-6.0) = $4.4^{+0.6}_{-0.5} \times 10^{35}$ ergs/s.  
\citet{Mukai00} observed XB 1832-330 with ASCA, finding that a partial covering model was required, instead of a single absorption column, to fit the spectrum.  A similar model was required by BeppoSAX \citep{Parmar01} and XMM \citep{Sidoli08} observations.  Our spectrum is of insufficient quality to distinguish between a partial covering model vs. a single absorption column.

We detect 9205 counts for source A in the 2000 HRC observation of NGC 6652. We convert the average source count rate 
into an unabsorbed bolometric X-ray luminosity with PIMMS using our best ACIS spectral fit,  
finding $L_X$(0.5-6.0) = (1.61 $\pm$ 0.02) $\times 10^{36}$ ergs/s. 
A's luminosity appears to have decreased by almost a factor of four from 2000 to 2008.  
\citet{Sidoli08} report two XMM observations in Sept. and Oct. 2006.  Converting their X-ray luminosities to the 0.5-6 keV range, they find $L_X$(0.5-6 keV)=$1.2\times 10^{36}$ and $1.1\times 10^{36}$ergs/s.   \citet{Parmar01} report a BeppoSAX measurement in March 2001, which we convert to $L_X$(0.5-6 keV)=$1.5\times 10^{36}$ ergs/s.  \citet{Tarana07} report 2003-2005 INTEGRAL measurements (restricted to $>$20 keV, and thus not directly comparable to our fluxes) that are consistent with the BeppoSAX flux. 
RXTE PCA Galactic Bulge Scan monitoring of X1832-330\footnote{http://lheawww.gsfc.nasa.gov/users/craigm/galscan/html/R\_1832-330.html} indicates substantial variability on $\sim$6-month time scales, but also a general declining trend from 1999 to 2010, by about 30\%.

\begin{figure}
\figurenum{2}
\includegraphics[angle=-90,scale=0.35]{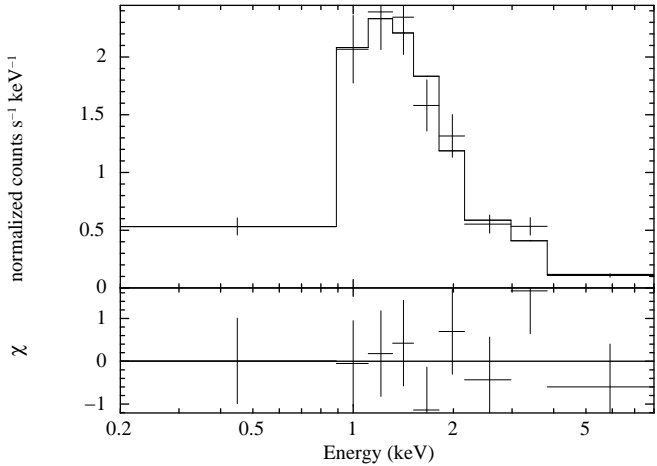}
\caption{Top: \textit{Chandra} 2008 X-ray spectrum of source A containing 9 bins of 50 counts per bin. The spectrum is best-fit by an absorbed power-law model of photon index of $\Gamma$ = $1.7^{+0.3}_{-0.2}$. Bottom: Residuals to the best fit.} 
\end{figure}

{\bf 3.2. Source B}

{\bf 3.2.1. Timing Analysis}

We produced a lightcurve by binning the 0.3-7 keV data into 103 bins each with 50s per bin (Fig. 3), which shows clear variability, by factors $>$10 on timescales $<$100 s.  (A K-S test indicates variability at $>$99.9\% confidence.)  Such variability is unusual in low-mass X-ray binaries.  To test whether the variation may be caused by changes in obscuring column (perhaps, if the system is edge-on, caused by material at the rim of an accretion disk), we group the lightcurve by countrate, with boundaries when the countrate crosses 0.15 and 0.3 counts/s (as the count statistics are low for 50 s bins).  Within each larger bin, we compute the hardness ratio of 0.3-1 keV photons (the most likely to be absorbed) over 0.3-7 keV photons (Fig. 3, bottom), with binomial 1$\sigma$ errorbars derived from \citet{Gehrels86}.  If the dips were due to obscuration, we might expect the lowest-countrate bins to have the lowest hardness ratios.  This is suggested by the first 1000 s, but is less clear from the dataset as a whole.  We quantify this by comparing the ratio of 0.3-1 keV counts to total counts in the low-countrate portions ($<$0.15 cts/s), vs. medium and high-countrate portions (below and above 0.3 cts/s), finding the fraction to be 0.15$\pm0.03$ for the low-countrate portions vs. 0.25$\pm0.04$ and 0.23$\pm0.03$ for the higher-countrate portions.  This is a significant effect, indicating that obscuration may play a role in the dipping.

 The peaks in the lightcurve reach countrates that suggest substantial pileup.  We estimate the amount of pileup and intrinsic luminosities using the PIMMS tool, for the best-fit absorbed power-law spectrum (see below; the choice of spectrum does not appreciably affect results).  At the peak observed countrate of 0.46 cts/s, the estimated fraction of recorded events that are actually multiple is 27\%, and we infer the intrinsic $L_X$(0.3-7)$\sim9\times10^{34}$ ergs/s, a factor of 2 higher than would be extrapolated without pileup.  However, the pileup model in PIMMS is systematically uncertain, so the true peak $L_X$ is uncertain, presumably by less than a factor of 2.  The inferred minimum $L_X$ is $<2\times10^{33}$ ergs/s, for 0.02 cts/s (cf. Fig. 3).

\begin{figure}[h]
\figurenum{3}
\includegraphics[angle=0,scale=0.45]{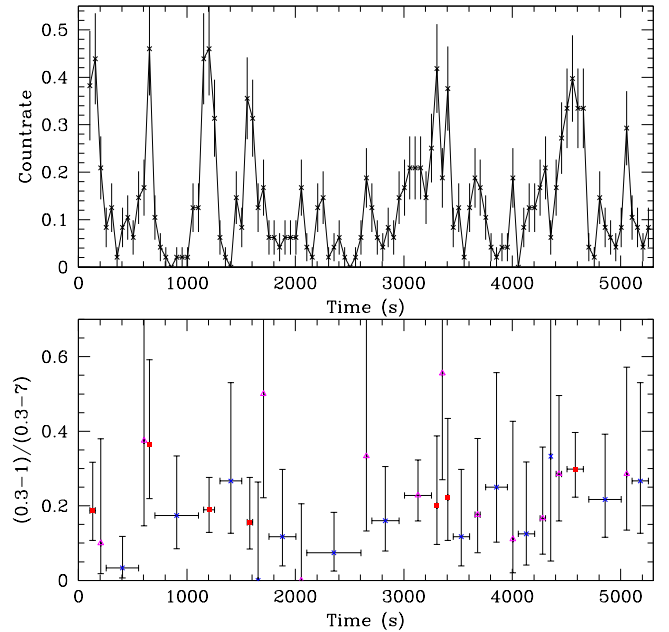}
\caption{Top: ACIS countrate lightcurve of source B, 0.3-7 keV, in 50 s bins.  $L_X$/ctrate conversion estimated as $1.2\times10^{34}$ ergs/s for 0.1 cts/s, reaching $9\times10^{34}$ ergs/s for 0.46 cts/s, due to pileup.  Bottom: Ratio of counts in 0.3-1 keV vs. those in 0.3-7 keV.  Boundaries of ratio bins set by when countrate crosses 0.15 and 0.3 keV.  Bins over 0.3 cts/s: (red) filled squares; 0.3-0.15 cts/s: (magenta) open triangles; and under 0.15 cts/s: (blue) crosses.
} 
\end{figure}

Power spectra (produced using XRONOS\footnote{http://heasarc.nasa.gov/docs/xanadu/xronos/xronos.html}) show no evidence of periodicity during the 5.6 ks ACIS observation. 
We re-examined the 2000 HRC observation to check for long-term variability.  Using PIMMS and our best ACIS spectral fit (below), we estimate the unabsorbed luminosity to be $L_X$(0.5-6.0) = (1.1 $\pm$ 0.1) $\times$ $10^{34}$ ergs/s, vs. our average ACIS unabsorbed luminosity of $L_X(0.5-6.0)$ = $1.1^{+0.1}_{-0.1}$ $\times$ $10^{34}$ ergs/s (section 3.2.2).  \citet{Heinke01} found significant variability during the HRC observation.  The short, low-sensitivity lightcurve suggests variation by at least a factor of two.  As the HRC sensitivity and exposure length are each 3 times smaller than those for the ACIS observation, its usefulness for studying variability is limited.

An archival ROSAT HRI observation (March 27, 1994, for 817 s exposure time) shows marginal evidence for source B.  The image (Fig. 4) shows one bright source, which we attribute to A, setting our astrometry.  We measure 6 photons within 2'' of source B's position, while only 1.4$^{+2.5}_{-1}$ photons are expected at this position, using the average background rate between 6.5'' and 10'' from source A, giving 4.6$^{+1}_{-2.5}$ (at 1 sigma; 4.6$^{+1.3}_{-4.6}$ at 2 sigma, \citealt{Gehrels86}) photons from B.  This $\sim$$2\sigma$ detection suggests (correcting for the HRI point-spread function enclosed energy of 38\% within 2") $L_X$(0.5-6 keV)$=1.5^{+0.3}_{-0.8}\times10^{34}$ ergs/s.  
Thus, we have no evidence for B's variability on long time scales.

\begin{figure}[h]
\figurenum{4}
\includegraphics[scale=0.45]{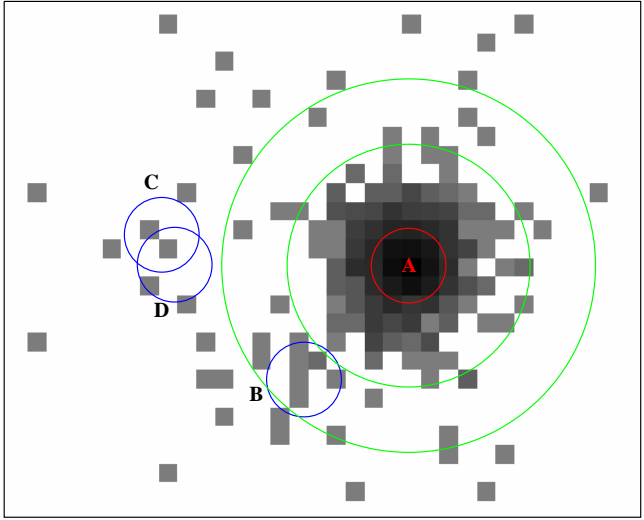}
\caption{ROSAT HRI image, March 1994, of NGC 6652.  The (corrected, using A's position) positions of sources A, B, C and D are marked with small (radius=2'') circles, while the annulus around A used to estimate the background is indicated with larger circles (radius 6.5'' and 10'').} 
\end{figure}

{\bf 3.2.2. Spectral Analysis}

We created both an overall time-averaged spectrum of B (680 counts, binned by 30 counts/bin), and spectra for the three countrate ranges listed above.
We include the Chandra CCD pileup model \citep{Davis01}, with the grade morphing parameter $\alpha$ fixed to 0.5, and photoelectric absorption (XSPEC {\it phabs}). 
B's time-averaged spectrum can be fit well with an absorbed power-law model with photon index $\Gamma$ = $1.24^{+0.10}_{-0.23}$ (Table 2). 
 We also try adding a NS hydrogen atmosphere model (NSATMOS, \citealt{Heinke06a}) to the absorbed power-law model.  We fix the mass and radius to canonical values (1.4 \Msun, 10 km), so kT is the only free parameter. 
The NSATMOS component does not improve the quality of the spectral fit, and the upper limit on its kT implies an upper limit on $L_{X,NS}$(0.5-10.0 keV) $< 7\times$ $10^{32}$ ergs/s. This limit may not be applicable if the system is edge-on, as the NS could suffer higher $N_H$ than other X-ray emitting regions.

We next fitted the three countrate-selected spectra, with 240, 186, and 253 counts [low to high], binned by 15 counts each, to investigate what parameters may be changing. 
An absorbed, piled-up power-law spectrum is best fit with (at least) two parameters varying; the power-law normalization, and either the $N_H$ or power-law spectral index. We give parameters for both such fits in Table 3.
The alternative of allowing only the power-law normalization to vary gives significant residuals in the spectra (Fig. 5), and a higher chi-squared: 42.93 for 38 degrees of freedom (dof), vs. 30.3 or 30.6 (for 36  dof) from the other two models (respectively).  An F-test gives probabilities of 0.2\% of attaining such an improvement in the chi-squared by chance, indicating that another quantity besides the normalization is also varying.

\begin{figure}
\figurenum{5}
\includegraphics[angle=-90,scale=0.37]{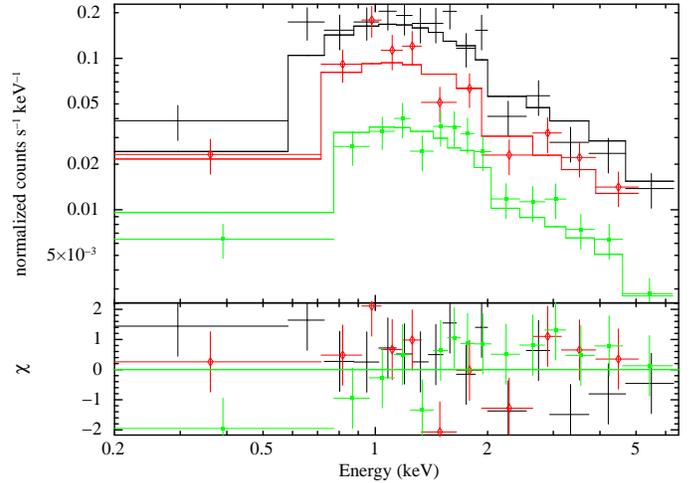}
\caption{Top: \textit{Chandra} X-ray spectra of source B at high (black, crosses), medium (red, diamonds) and low (green, filled squares) countrates.  All are fit with a power-law model with only normalization varying (see text). Bottom: Residuals to this fit, demonstrating the spectral changes.} 
\end{figure}

{\bf 3.3. Source C:}

An unbinned 0.3-7 keV lightcurve gives a 0.01 K-S probability of source constancy, indicating that C is variable. This is confirmed by its (0.3-7 keV, 500 s binning) lightcurve, which shows two clear peaks (Fig. 6). A power spectrum shows no evidence of periodicity. 
Using the best-fit double mekal spectral model (below), we find a peak $L_X$(0.5-10.0) = $3.4^{+1.1}_{-1.1}$ $\times$ $10^{33}$ ergs/s, and a minimum $L_X< 3.2 \times$ $10^{32}$ ergs/s, for an average $L_X=1.1\times10^{33}$.  Using the power-law spectral model (a poor fit) to the ACIS data to infer the HRC spectrum, we find $L_X=2\times10^{32}$ ergs/s, half the ACIS estimate with this spectrum, but the spectral uncertainties render this conclusion uncertain.





\begin{figure}
\figurenum{6}
\includegraphics[angle=0,scale=0.6]{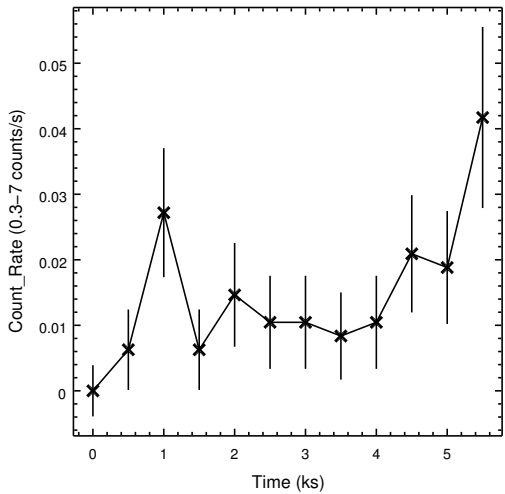}
\caption{ \textit{Chandra} lightcurve of source C, 0.3-7 keV, at 500 s binning.  } 
\end{figure}

We group C's spectrum by 10 counts/bin for $\chi^2$ statistics (Fig. 7), and also fit the unbinned spectrum with C-statistics (finding similar results). No single model we tried can fit this very soft spectrum, with reduced $\chi^2$ values well above 2 (Table 2). 
The spectrum can be fit by a model containing two MEKAL \citep[thermal plasma,][]{Liedahl95} components with the cluster absorption (to simplify fitting we fixed $N_H$), with temperatures of $<0.096$ and $>2.3$ keV. 
Although a double MEKAL model is often used to describe the spectra of CVs \citep[e.g.][]{Baskill05}, such a strong low-temperature component is rarely seen.  We are aware of such components only in nova remnants \citep{Balman05}. 
Alternatively, an absorbed blackbody and MEKAL represents a simplified form of a model for polar CVs \citep[e.g.][]{Ramsay04a}. 
C's spectrum is reasonably fit by a blackbody plus MEKAL model, with a blackbody temperature of $67^{+14}_{-14}$ eV and inferred radius of $46^{+58}_{-23}$ km, and a MEKAL temperature of $>$3.2 keV.  

\begin{figure}
\figurenum{7}
\includegraphics[angle=-90,scale=0.33]{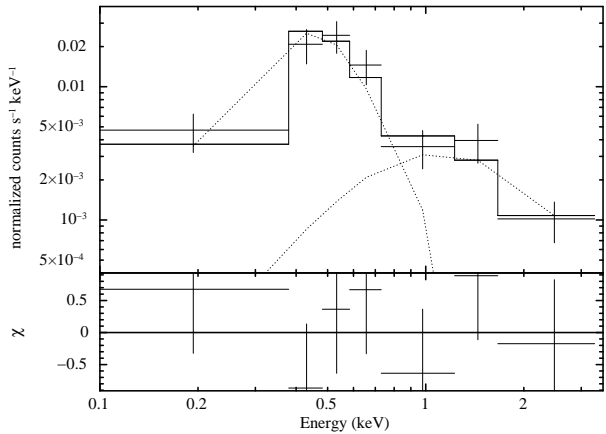}
\caption{Top: \textit{Chandra} 2008 X-ray spectrum of source C, fit by a low-temperature blackbody plus MEKAL model.  Dotted lines indicate the two components (the blackbody is the lower-temperature component).  Bottom: Residuals to the best fit.} 
\end{figure}

{\bf 3.4. Source D:}

A K-S test on the unbinned lightcurve gives a probability of constancy of 0.46. The estimated luminosity from the HRC data using the ACIS power-law fit below is $L_X$(0.5-10.0) = 1.1$\times$ $10^{33}$ ergs/s. This is comparable to the luminosity from the ACIS data, $L_X$(0.5-10.0) = $8^{+5}_{-3}$ $\times$ $10^{32}$ ergs/s, so no variability is apparent.

We extract a 69-count spectrum, binned with 10 counts/bin.
Source D is relatively soft, with all 6 bins below 2.0 keV (Fig. 8), motivating us to freeze $N_H$ to the cluster value. An NSATMOS model is a poor fit, but an absorbed powerlaw (photon index of $2.3^{+0.6}_{-0.6}$) is an adequate fit to the binned spectrum; a single-temperature MEKAL is a somewhat worse fit (Table 2; null hypothesis probabilities of 25\% and 11\% respectively).  The powerlaw fit, however, shows residuals suggesting strong line emission around 1 keV (Fig. 8). Adding a gaussian line of zero width to the power-law fit gives a line energy of 1.03$^{+0.12}_{-0.04}$ keV, but does not improve the fit by a statistically significant amount (an F-test indicates a probability of 43\% that such an improvement could happen by chance).  Fitting the unbinned data (using the C-statistic) with a power-law gives a slightly larger photon index ($2.7^{+0.4}_{-0.4}$), and the fraction of simulated spectra with lower C-statistic values is 97\%, suggesting a relatively poor fit. (The single-temp MEKAL fit to the unbinned spectrum gives 96\% of simulations with lower C-statistic values, as well.)   

\begin{figure}
\figurenum{8}
\includegraphics[angle=-90,scale=0.33]{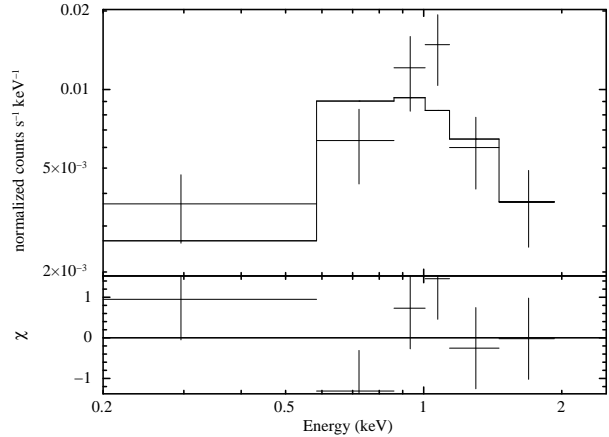}
\caption{Top: \textit{Chandra} spectrum of source D, fit to an absorbed power-law model (see text). Bottom: Residuals to the best fit.} 
\end{figure}

{\bf 3.5. Faint Sources:}

We detect only 15, 7, and 15 events for source E, F, and G respectively, too few for detailed spectral and timing analyses.  
We extract background-subtracted count rates in energy bands of 0.5-6.0 keV, 0.5-1.5 keV, and 1.5-6 keV, using the WAVDETECT regions and PIMMS to estimate unabsorbed source luminosities. 
While sources E and F have roughly equal numbers of soft and hard counts, all but one of source G's counts lie above 1.5 keV, suggesting strong absorption.  

Using PIMMs, we estimate the unabsorbed luminosities of the faint sources in both the 0.5-1.5 keV and 1.5-6.0 keV energy bands, using an absorbed power-law of photon index 1.4. We determine the total unabsorbed luminosity from 0.5 to 6.0 keV for each source by adding the soft and hard luminosities (Table 1).  We extract the counts at the ACIS WAVDETECT coordinates for each source from the HRC dataset, finding 4, 1, and 0 counts from E, F, and G respectively.  These upper limits are consistent with their measured ACIS luminosities.

{\bf 3.6. X-ray Color-Magnitude Diagram:}

We create an X-ray color-magnitude diagram (Fig. 9) containing the sources observed in the 2008 ACIS data, by plotting the unabsorbed 0.5-6.0 keV luminosities versus the X-ray color, defined as 2.5 log [(0.5-1.5 kev counts)/(1.5-6.0 keV counts)] \citep{Grindlay01b}.  
We plot an X-ray color-luminosity relation for the NSATMOS model, and X-ray color predictions for power-law and MEKAL spectral models. The spectra of qLMXBs are frequently modeled with a neutron star atmosphere \citep[e.g., NSATMOS][]{Heinke06b} component, with a harder power-law component with a photon index of 1-2 \citep{Rutledge02b}, making up anywhere from $<10$ to $>90$\% of the 0.5-10 keV flux \citep{Jonker04b}.  
We plot a representative NSATMOS plus power-law model in which the power-law component contributes 50\% of the 0.5-6.0 keV luminosity.  Note that B's position is shifted to the left by pileup (considering its rapid variability, it is hard to correct for this).

No sources lie near the NS atmosphere cooling track, but sources C and D have colors and luminosities in the range of qLMXBs in other clusters \citep[e.g.][]{Pooley06,Heinke06b}.  C's rapid variability and unusual spectrum likely rule out a qLMXB nature for it, but D could be a qLMXB with a dominant power-law spectral component. 

Surveys of X-ray sources in globular clusters indicate that most nonmagnetic CVs display hard spectra, typically consistent with MEKAL temperatures $>$ 6 keV or colors $<$ 0.5 \citep{Grindlay01b, Pooley02a}.
Sources E, F and G are consistent with the X-ray luminosities and spectra of CVs observed in many other globular clusters \citep{Pooley06,Heinke05a}, though quiescent LMXBs, active binaries, and millisecond pulsars cannot be ruled out. 
 G's extremely hard spectrum, though based on few counts, indicates a high intrinsic absorption ($>10^{22}$ cm$^{-2}$), and thus an edge-on CV \citep[e.g. W8, W15, W33, and AKO9 in 47 Tuc,][]{Heinke05a} or a background AGN.

\begin{figure}
\figurenum{9}
\includegraphics[scale=0.45]{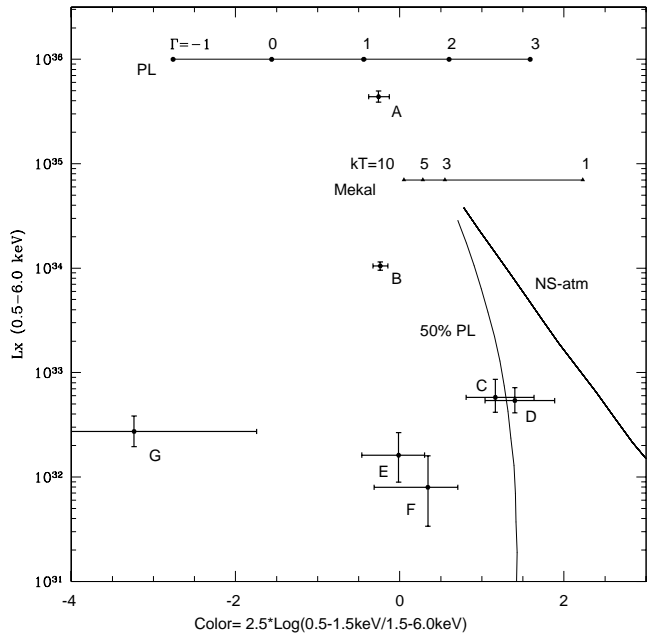}
\caption{X-ray color-magnitude diagram for all sources detected in the 2008 \textit{Chandra} ACIS-S observation. The color is defined as a function of the ratio of low-energy counts to high-energy counts. X-ray luminosities between 0.5 keV and 6.0 keV are plotted versus the color along with their respective errors. Also shown are the theoretical cooling tracks for the power-law, MEKAL, NSATMOS, and NSATMOS+power-law models. The NSATMOS+power-law model is defined such that 50\% of the model's 0.5-6 keV luminosity is produced by the power-law component.} 
\end{figure}

\section{Discussion}\label{s:discuss}

{\bf 4.1. Source B: Unusual LMXB}

Source B, one of the brightest low-$L_X$ cluster sources, can be classified as a very faint X-ray transient (VFXT). VFXTs are X-ray transients that have peak luminosities of $10^{34-36}$ ergs/s and quiescent luminosities at least one order of magnitude lower \citep{Wijnands06,Muno05b}, generally containing an accreting neutron star or black hole.  Some VFXTs have ``normal'' outbursts as well as very faint outbursts, but it is not clear if all do, or if VFXT behavior has a variety of causes \citep{Degenaar10}. 

Fig. 3 illustrates that source B reaches peak X-ray luminosities up to $L_X$(0.5-10.0 keV) = $9\times10^{34}$ ergs/s and minimum $L_X< 2\times10^{33}$ ergs/s on timescales of minutes. 
B's high peak luminosity is strong evidence that the system must contain a neutron star or a black hole, but the variability is unusual for LMXBs. 
Perhaps the simplest explanation for this variability is a high inclination angle, so that we observe B's accretion disk edge-on, and our view of the central X-ray source is interrupted by structures at the accretion disk rim \citep{White82,Xiang09}. 
Obscuration by an accretion disk should lead to changes in the $N_H$ value, and our spectral analysis gives evidence in favor of this (section 3.2.2).  Our spectral fitting requires intrinsic $L_X$ changes along with $N_H$ changes, which seems to argue against this explanation.  However, our spectral fitting is constrained by a lack of data to only three countrate ranges, which contain substantial variability, and we think it likely that variability also occurs on timescales shorter than we can accurately probe.  These two effects could prevent us from accurately measuring how $N_H$ changes with $L_X$ in B's lightcurve, and thus we cannot yet rule out that obscuration is fully responsible for the variability.

If we only see scattered light from a central source (in this picture the variability is due to obscuration of an accretion disk corona), the true isotropic X-ray luminosity could be  higher than the observed luminosity, by a factor of up to 100 \citep[e.g.][]{Muno05a}. Thus, B could be a 'normal' LMXB, which would remove the difficulty of explaining its unusually low accretion luminosity.  We have been awarded a deeper \Chandra\ observation of NGC 6652 (50 ks, to be taken mid-2011), which should clarify the spectral variability of this source.  We have also been awarded a Gemini time-series imaging observation of NGC 6652, designed to search for evidence of short periods in sources A and B suggested by previous HST imaging \citep{Deutsch00,Heinke01}.

{\bf 4.2. Source C: A Second Polar in a Globular Cluster?}

The very soft spectral component of C is unlike anything seen in quiescent LMXBs, but is consistent with the spectra of some CVs. The (low-T) blackbody + (high-T) MEKAL spectral fit suggests a polar CV nature, where the blackbody describes soft X-rays emitted from the white dwarf's polar cap, and the MEKAL describes hard X-rays from the accretion column shock front \citep{RamsayCropper04}. The inferred polar cap radius of $25^{+40}_{-14}$ km (from the blackbody spectral fit) compares well with the typical radius of an accreting pole of a WD of $\sim$75 km \citep[for a 0.6 \Msun WD,][]{Ishida97}. 
The spectrum and luminosity are similar to the likely polar CV X10  in the globular cluster 47 Tuc \citep{Heinke05a}, which has $L_X$(0.5-6 keV)$=2.6\times10^{32}$ ergs/s, and was modeled with a $kT=53$ eV blackbody and two MEKAL plasmas of temperatures 0.39 and $>$14 keV. 
Thus we suggest source C may be the second polar CV identified in a globular cluster.  Knowing the frequency of magnetic accretion channelling in cluster CVs (by identifying magnetic CVs) is important for understanding their unusually rare outbursts \citep{Grindlay99,Dobrotka06,Ivanova06}.
 
Alternatively, a very low-T thermal plasma has been seen in historical novae, from the nova shell \citep[e.g. ][]{Balman05}. However, the nova eruption would probably have been seen if recent, or if old, the shell should have expanded to a resolvable size (e.g., the Nova Per 1901 shell would be 2.5'' in radius if located in NGC 6652). Our upcoming \Chandra\ observation should provide sufficiently high-quality spectra to distinguish between the spectral models discussed, and search for spectral variations over time.

{\bf 4.3. Source D: unusual CV or quiescent LMXB?}

Source D's luminosity and spectrum are consistent with both CVs or quiescent LMXBs. The feature of particular interest is the hint of an emission line around 1 keV. If this feature is real (which will be tested by our upcoming \Chandra\ observation), it would make D a rather odd X-ray source. There is another X-ray bright CV with a similarly strong low-energy line, the luminous CV X9 in 47 Tuc \citep{Heinke05a}.  

{\bf 4.4. Faint Sources}

The pronounced spectral hardness of G indicates it is an edge-on CV, or a background AGN. E and F cannot be clearly classified, but as the majority of globular cluster X-ray sources with their X-ray colors and $L_X$ are CVs \citep{Pooley06}, we may suggest that CVs are the most likely possibility.  Deeper \Chandra\ observations should detect additional sources, enabling a comparison of the amazingly rich X-ray population of this globular cluster with other globular clusters.

\acknowledgements

We acknowledge support from NASA \Chandra\ grants, an NSERC Discovery Grant, and an Alberta Ingenuity New Faculty Award.  We thank M. Muno for his assistance in obtaining these observations.

\bibliography{src_ref_list}
\bibliographystyle{apj}


\begin{deluxetable}{lccccccccc}
\tablecolumns{10}
\tabletypesize{\scriptsize}
\tablecaption{\textbf{Sources in NGC 6652 Field}}
\tablehead{
\multicolumn{2}{c}{\textsc{Source}} & & & & & & \multicolumn{3}{c}{$L_X$ ($10^{32}$ ergs s$^{-1}$)}  \\ 
\colhead{Name} & \colhead{CXOGLB J} & \colhead{\textsc{R.A.}} & \colhead{\textsc{Err}} & \colhead{\textsc{Decl.}} & \colhead{\textsc{Err}} & \textsc{Counts} & & &  \\
& & & & & & (0-10 keV) & (0.5-1.5 keV) & (1.5-6 keV) & (0.5-6 keV) \\
}
\startdata
A  & 183543.7-325927 & 18:35:43.69 & 0.02 & -32:59:26.5 & 0.5 & 462$^a$ & $1574^{+533}_{-386}$ & $2813^{+312}_{-310}$ & $4386^{+599}_{-507}$ \\
B  & 183544.6-325938 & 18:35:44.567  & 0.001 & -32:59:38.36 & 0.02 & 680 & $27^{+7}_{-5}$ & 77 $\pm$ 8 & $105^{+10}_{-9}$ \\
C  & 183545.8-325923 &18:35:45.755  & 0.003 & -32:59:23.21 & 0.04 & 90 & $3.9^{+0.9}_{-0.9}$ & $1.9^{+2.6}_{-1.1}$ &$5.8^{+2.8}_{-1.6}$ \\
D  & 183545.6-325926 & 18:35:45.648  & 0.004 & -32:59:26.10 & 0.05 & 73 & $3.3^{+0.8}_{-0.7}$ & $2.1^{+1.8}_{-0.9}$ & $5.4^{+1.8}_{-1.3}$ \\
E  & 183546.2-325929 & 18:35:46.206 &0.010 &-32:59:29.26 & 0.09 & 15 & $0.47^{+0.28}_{-0.19}$  & $1.2^{+0.8}_{-0.5}$  & $1.6^{+1.0}_{-0.7}$ \\
F & 183547.8-325920 &18:35:47.777 & 0.004 & -32:59:19.84 & 0.15 & 7 & $0.27^{+0.23}_{-0.14}$  & $0.53^{+0.57}_{-0.32}$  & $0.80^{+0.80}_{-0.46}$ \\
G  & 183548.8-325921 & 18:35:48.816 & 0.009 &-32:59:21.13 & 0.10 & 15 & $0.05^{+0.17}_{-0.05}$  & $2.7^{+0.9}_{-0.7}$  & $2.7^{+1.1}_{-0.8}$ \\
\enddata
\tablecomments{Positions (with relative errors, not including systematic uncertainties on the \Chandra\ astrometry), total counts, and inferred luminosities for NGC 6652 sources.  \\
$^a$: From readout streak.
}
\end{deluxetable}

\begin{deluxetable}{lcccccc}
\tablewidth{15cm}
\tabletypesize{\small}
\tablecaption{\textbf{Spectral Fits}}
\tablehead{
{\textbf{Src}} & {\textbf{Model}}  & \textbf{$N_{H}$} & \textbf{$\Gamma$} & \textbf{kT} & \textbf{$\chi^{2}_{\upsilon}$/dof} & \textbf{$L_X$}  \\
 & & ($10^{20}$ cm$^{-2}$) & & keV & 
}
\startdata
A & POW & $27^{+10}_{-8}$ & $1.7^{+0.3}_{-0.2}$ & - & 0.87/6 & $5.8^{+0.8}_{-0.7}\times10^{35}$  \\
\hline
B  & POW & $5^{+6}_{-0}$ & 1.3$^{+0.2}_{-0.2}$ & - & 0.70/18 & 1.7$^{+0.2}_{-0.2}\times10^{34}$  \\
B  & POW+NSATMOS & $5^{+8}_{-0}$ & 1.2$^{+0.2}_{-0.2}$ & - & 0.736/17 & $1.7^{+0.2}_{-0.2}\times10^{34}$ \\
- & NSATMOS & - & - & $10^{+100}_{-10}$ & - & $<7\times10^{32}$ \\
\hline
C & POW & (5) & 5.2 & - & 3.14/5 & 4.1$\times10^{32}$ \\
C & MEKAL & (5) & - & $0.14$ & 6.24/5 & 5.6$\times10^{32}$  \\
C & MEKAL+MEKAL & (5) & - & $0.08^{+0.015}_{-0}$ & 0.45/3 & 1.1$^{+0.9}_{-0.5}\times10^{33}$  \\
 - & 2nd MEKAL & - & - & $>2$ &- & $7.3^{+2.8}_{-4.3}\times10^{32}$  \\
C & MEKAL+BBODY & (5) & - & $>3.3$ & 1.00/3 & 1.2$^{+0.1}_{-0.5}\times10^{33}$  \\
 -  & BBODY & - & - & $0.067^{+0.013}_{-0.013}$ & - & $3.1^{+1.2}_{-1.3}\times10^{32}$ \\
\hline
D & POW & (5) & $2.3^{+0.6}_{-0.6}$ & - & 1.33/4 & 8$^{+5}_{-3}\times10^{32}$  \\
D & MEKAL & (5) & - & $2^{+5}_{-1}$ & 1.88/4 & 6.8$^{+6.2}_{-2.5}\times10^{32}$ \\
D & NSATMOS & (5) & - & 0.1 & 2.3/5 & $4.3\times10^{32}$ \\
D & POW+GAU & (5) & $2.3^{+0.7}_{-0.7}$ & - & 1.14/2 & 8$^{+5}_{-3}\times10^{32}$  \\
- & GAU & - & 1.03$^{+0.12}_{-0.04}$ & - & - & $5^{+5}_{-5}\times10^{31}$ \\
\enddata
\tablecomments{Spectral fits to NGC 6652 sources (see text for details). Errors are 90\% confidence for a single parameter, and are not computed if $\chi^2_{\nu}>2$.  Luminosities are unabsorbed in erg s$^{-1}$ for 0.5-10 keV.  Two-component models are continued on a second line (omitting the source name in the first column), with the total luminosity on the first line and the second component's luminosity individually on the second line. All models include PHABS; for faint, soft sources, we have fixed the $N_H$ to the cluster value. 
}
\end{deluxetable}

\begin{deluxetable}{lccccc}
\tabletypesize{\small}
\tablewidth{12cm}
\tablecaption{\textbf{Countrate-Resolved Spectral Fits to B}}
\tablehead{
{\textbf{Source}} & {\textbf{Model}}  & \textbf{$N_{H}$} & \textbf{$\Gamma$} &  \textbf{$\chi^{2}_{\upsilon}$/dof} & \textbf{$L_X$}  \\
& & $(10^{20} cm^{-2})$ &  & & ergs/s 
}
\startdata
\multicolumn{6}{c}{Allowing PL norm and index to vary}\\
\hline
B (High) & PHABS(POW) & (5) & $1.5^{+0.2}_{-0.2}$ &  0.84/36 & 5.1$^{+1.1}_{-1.0}\times10^{34}$  \\
B (Med) & ... & ... & $1.3^{+0.3}_{-0.3}$ & ... &  2.7$^{+0.7}_{-0.7}\times10^{34}$ \\
B (Low)  & ... & ... & $1.0^{+0.2}_{-0.2}$ & ... &  1.2$^{+0.2}_{-0.2}\times10^{34}$  \\
\hline
\multicolumn{6}{c}{Allowing PL norm and $N_H$ to vary}\\
\hline
B (High) & PHABS(POW) & $5.0^{+3.3}_{-0}$ & $1.4^{+0.7}_{-0.4}$ &  0.83/37 & 5.5$^{+1.1}_{-1.0}\times10^{34}$  \\
B (Med) & ... & $5.9^{+7.6}_{-0.9}$ & ... & ... &  $2.4^{+0.5}_{-0.5}\times10^{34}$ \\
B (Low)  & ... & $22^{+12}_{-9}$ & ... & ... &  $1.1^{+0.2}_{-0.2}\times10^{34}$  
\enddata
\tablecomments{Spectral fits to NGC 6652 B, split by countrate (see text). Errors are 90\% confidence for a single parameter.  Luminosities are unabsorbed in erg s$^{-1}$ for 0.5-10 keV.
}
\end{deluxetable}

\end{document}